\documentclass[twocolumn,twoside,prl,superscriptaddress]{revtex4}
\usepackage{amsmath}
\usepackage{amssymb}
\usepackage{epsfig}

\newcommand{\ket}[1]{| #1 \rangle}

\newcommand{\tr}{\operatorname{tr}}
\newcommand{\av}[1]{\left\langle #1 \right\rangle_t}
\newcommand{\eff}{\rm eff}
\newcommand{\nn}{\nonumber}
\newcommand\ra{\rangle}
\newcommand\la{\langle}

\def\squareforqed{\hbox{\rlap{$\sqcap$}$\sqcup$}}
\def\qed{\ifmmode\squareforqed\else{\unskip\nobreak\hfil
\penalty50\hskip1em\null\nobreak\hfil\squareforqed
\parfillskip=0pt\finalhyphendemerits=0\endgraf}\fi}
\def\endenv{\ifmmode\;\else{\unskip\nobreak\hfil
\penalty50\hskip1em\null\nobreak\hfil\;
\parfillskip=0pt\finalhyphendemerits=0\endgraf}\fi}

\mathchardef\ordinarycolon\mathcode`\:
\mathcode`\:=\string"8000
\def\vcentcolon{\mathrel{\mathop\ordinarycolon}}
\begingroup \catcode`\:=\active
  \lowercase{\endgroup
  \let :\vcentcolon
  }

\newcommand{\nc}{\newcommand}
\nc{\rnc}{\renewcommand}
\nc{\beq}{\begin{equation}}
\nc{\eeq}{{\end{equation}}}
\nc{\beqa}{\begin{eqnarray}}
\nc{\eeqa}{\end{eqnarray}}
\nc{\lbar}[1]{\overline{#1}}
\nc{\ketbra}[2]{|#1\rangle\!\langle#2|}
\nc{\avg}[1]{\langle#1\rangle}
\rnc{\max}{\operatorname{max}}
\nc{\Rank}{\operatorname{Rank}}
\nc{\smfrac}[2]{\mbox{$\frac{#1}{#2}$}}
\nc{\ox}{\otimes}
\nc{\dg}{\dagger}
\nc{\dn}{\downarrow}
\nc{\cA}{\mathcal{A}}
\nc{\cB}{\mathcal{B}}
\nc{\cC}{\mathcal{C}}
\nc{\cD}{\mathcal{D}}
\nc{\cE}{\mathcal{E}}
\nc{\cF}{\mathcal{F}}
\nc{\cG}{\mathcal{G}}
\nc{\cH}{\mathcal{H}}
\nc{\cI}{\mathcal{I}}
\nc{\cJ}{\mathcal{J}}
\nc{\cK}{\mathcal{K}}
\nc{\cL}{\mathcal{L}}
\nc{\cM}{\mathcal{M}}
\nc{\cN}{\mathcal{N}}
\nc{\cO}{\mathcal{O}}
\nc{\cP}{\mathcal{P}}
\nc{\cR}{\mathcal{R}}
\nc{\cS}{\mathcal{S}}
\nc{\cT}{\mathcal{T}}
\nc{\cX}{\mathcal{X}}
\nc{\cZ}{\mathcal{Z}}
\nc{\csupp}{{\operatorname{csupp}}}
\nc{\qsupp}{{\operatorname{qsupp}}}
\nc{\var}{\operatorname{var}}
\nc{\rar}{\rightarrow}
\nc{\lrar}{\longrightarrow}
\nc{\polylog}{\operatorname{polylog}}
\nc{\1}{{\openone}}
\nc{\id}{{\operatorname{id}}}

\nc{\RR}{{{\mathbb R}}}
\nc{\CC}{{{\mathbb C}}}
\nc{\FF}{{{\mathbb F}}}
\nc{\NN}{{{\mathbb N}}}
\nc{\ZZ}{{{\mathbb Z}}}
\nc{\PP}{{{\mathbb P}}}
\nc{\QQ}{{{\mathbb Q}}}
\nc{\UU}{{{\mathbb U}}}
\nc{\EE}{{{\mathbb E}}}

\nc{\be}{\begin{equation}}
\nc{\ee}{{\end{equation}}}
\nc{\bea}{\begin{eqnarray}}
\nc{\eea}{\end{eqnarray}}
\nc{\<}{\langle}
\nc{\Hom}[2]{\mbox{Hom}(\CC^{#1},\CC^{#2})}
\nc{\rU}{\mbox{U}}

\begin{document}

\title{On the speed of fluctuations around thermodynamic equilibrium}

\author{Noah Linden}
\affiliation{Department of Mathematics, University of Bristol,
University Walk, Bristol BS8 1TW, U.K.}

\author{Sandu Popescu}
\affiliation{H.H.Wills Physics Laboratory, University of Bristol,
Tyndall Avenue, Bristol BS8 1TL, U.K.}

\author{Anthony J. Short}
\affiliation{Department of Applied Mathematics and Theoretical
Physics, University of Cambridge, Centre for Mathematical Sciences,
Wilberforce Road, Cambridge CB3 0WA, U.K.}

\author{Andreas Winter}
\affiliation{Department of Mathematics, University of Bristol,
University Walk, Bristol BS8 1TW, U.K.} \affiliation{Centre for
Quantum Technologies, National University of Singapore, 2 Science
Drive 3, Singapore 117542}

\date{}

\begin{abstract}

We study the speed of fluctuation of a quantum system around its
thermodynamic equilibrium state, and show that the speed will be extremely small
for almost all times in typical thermodynamic cases. The setting considered here is that of a quantum system couples to a bath, both jointly
described as a closed system. This setting, is the same as the one considered in [N. Linden \emph{et al.}, Phys. Rev. E 79:061103 (2009)]
and the ``thermodynamic equilibrium state'' refers to a situation that includes the usual thermodynamic equilibrium case, as well as far more general situations.
\end{abstract}

\maketitle

Recently there has been significant progress in understanding the fundamental principles of statistical
mechanics~\cite{Mahler:etal, Lebowitz, PopescuShortWinter, bocchieri:loinger, lloyd, Tasaki, reimann,
goldstein}. Underlying this progress is the realization that quantum mechanics allows individual quantum states
to exhibit statistical properties, and that ensemble or time averages are not needed to obtain a mixed
equilibrium state for the system under consideration. This is a purely quantum phenomenon, and the key is
entanglement, which leads to \emph{objective} uncertainty -- even when we have complete knowledge of the state
of the whole system, a subsystem that is entangled with the rest of the system will be best described by a mixed
state (i.e., a probabilistic mixture of pure states).

This progress led to a proof from first principles that virtually any subsystem of any large enough system will
reach equilibrium and fluctuate around it for almost all times \cite{linden}.
In order to better understand the nature of these
fluctuations, and to help understand the time-scales involved,
here we investigate the speed of fluctuations around
equilibrium: does the state of the subsystem oscillate rapidly
around equilibrium, or remain relatively static?  Our main result is
to put a universal upper bound on the average speed of fluctuations,
showing that the speed is extremely small for almost all times in
typical thermodynamic cases.

\medskip
Consider a large quantum system, described by a Hilbert space
$\mathcal{H}$. We decompose this system into two parts, a small
subsystem $S$, and the rest of the system that we refer to as the
bath $B$. Correspondingly, we decompose the total Hilbert space as
$\mathcal{H} =\mathcal{H}_S \otimes \mathcal{H}_B$, where
$\mathcal{H}_S$ and $\mathcal{H}_B$ (of dimensions $d_S$ and $d_B$)
are the Hilbert spaces of the subsystem and bath respectively. If
the subsystem or bath would have infinite dimension, we bound its
volume and energy to render the dimension finite, and project the
interaction Hamiltonian onto the restricted Hilbert space
$\mathcal{H}$.

Let the subsystem and bath evolve under a Hamiltonian $H$ that we
decompose into a constant, subsystem, bath and interaction term
\begin{equation}
  H= H_0 +  H_S  +  H_B + H_{\rm int}
\end{equation}
The decomposition is made unique by taking $H_0$ proportional to the
identity, $H_{S}$ the tensor product of a traceless operator on the
subsystem and the identity on the bath, $H_{B}$ the tensor product of
the identity on the subsystem and a traceless operator on the bath,
and $H_{\rm int}$ traceless for both the subsystem and bath.
The total system, i.e. subsystem plus bath, is in a pure state
$|\Psi(t)\ra$;  let $\rho(t)=|\Psi(t)\ra\!\la\Psi(t)|$ be the density
matrix representation of the state of the total system and let
$\rho_S(t)=\tr_B \rho(t)$ be the density matrix of the subsystem.

Following the notation in \cite{linden} we  define the time-averaged
state of the whole system $\omega$, which is given by
\begin{equation} \label{eqn:notation1}
  \omega = \av{\rho(t)} = \lim_{\tau \rightarrow \infty} \frac{1}{\tau} \int_0^{\tau} \rho(t) {\rm d}t.
\end{equation}
Similarly we define $\omega_S = \tr_B \omega$ and
$\omega_B  = \tr_S \omega$ as the time-averaged
states of the subsystem and bath respectively.
It is also convenient to introduce the notion of the \emph{effective
dimension} of a (mixed) state $\rho$ by
\begin{equation} \label{eqn:notation2}
  d^{\eff} (\rho) = \frac{1}{\tr(\rho^2)}.
\end{equation}
This is a measure of the number of states over which $\rho$ is
spread (e.g. for an equal mixture of $N$ orthogonal states,
$d^{\eff}=N$).

Clearly, $\omega_S$ is the canonical candidate for the equilibrium state, but we need
to pause to clarify just what it may mean that the system
reaches equilibrium. Namely, note that due to the finiteness of the
 Hilbert spaces involved, there will be recurrences [on timescales
exponential in $d^{\eff}(\omega)$], so a relaxation of $\rho_S(t)$
towards $\omega_S$ is out of the question. The best thing we can
hope for in the current setting is that $\rho_S(t)$ remains close to
$\omega_S$ for most times $t$.

To put our present results in a proper context, we first recall the key ideas and results of   \cite{linden}.  A key observation was that
 the process
of thermalisation actually contains many different aspects and we can decompose
it into the following elements of analysis.

{\it 1. Equilibration}. We say that a system equilibrates if its state evolves towards some
particular state (in general mixed) and remains in that state (or close to it) for almost all times. As far as
equilibration is concerned, it is irrelevant what the equilibrium state actually is.

{\it 2. Bath state independence}. The equilibrium state of the system should not
depend on the precise initial state of the bath, but only on its macroscopic parameters (e.g. its temperature)

{\it 3. Subsystem state independence}. If the subsystem is small
compared to the bath, the equilibrium state of the subsystem should be independent of its initial state.

{\it 4. Boltzmann form of the equilibrium state.} Under certain
additional conditions on the Hamiltonian (especially the interaction term) and on the initial state, the
equilibrium state of the subsystem can be written in the familiar Boltzmannian form $\rho_S =
\frac{1}{Z}\exp\left(-\frac{\widetilde{H}_S}{k_B T}\right)$.

Realizing that thermalization can be decomposed in this way had major consequences.
First, it allowed us to address each aspect separately. Second, and more important,
it allowed us to greatly expand the scope of our study. Indeed, we now consider
equilibration as a general quantum phenomenon that may occur in situations other
than those usually associated with thermalization. In particular we need not
restrict ourselves to any of the following: standard
thermal baths (that are described by a given temperature or restricted energy range),
weak or short range interactions between the system and the bath, Boltzmannian
distributions, situations in which energy is an extensive quantity, etc.
Furthermore, we can consider situations in which the subsystem does not reach equilibrium,
and prove results about the bath or subsystem independence properties of the time-averaged state.

In \cite{linden}, we made substantial progress on items {\it 1, 2}, and {\it 3} above,
under very weak assumptions. The only real constraint we impose
on the Hamiltonian is that it has non-degenerate energy gaps. That is, any four energy
eigenvalues satisfy $ E_1 - E_2 = E_3 - E_4$ if and only if $E_1=E_2$ and $E_3=E_4$,
or $E_1=E_3$ and $E_2=E_4$. This assumption rules out non-interacting Hamiltonians of
the form $H_S \neq 0, H_B \neq 0, H_{\rm int} = 0$, which obviously do not equilibrate.
These conditions are essentially those in \cite{linden} but allow greater flexibility. Indeed, they
allow the Hamiltonian to have degenerate energy levels, as long as the gaps between levels are non-degenerate, while in \cite{linden} degenerate levels were not allowed. Yet, it can be easily shown that even under these more general conditions the results of \cite{linden} hold, namely that any small subsystem will
reach equilibrium and fluctuate around it for almost all times \cite{degenerate}.

More precisely, \cite[Theorem 1]{linden} shows that the average
distance between $\rho_S(t)$ and its time average
$\omega_S$ is bounded by
\begin{equation} \label{prev_result}
    \av{ D( \rho_S(t), \omega_S )}
       \leq \frac{1}{2} \sqrt{\frac{d_S}{d^{\eff} (\omega_B)}}
       \leq \frac{1}{2} \sqrt{\frac{d_S^2}{d^{\eff} (\omega)}},
  \end{equation}
where $D(\rho_1, \rho_2 )=\frac{1}{2} \| \rho_1 - \rho_2 \|_1
=\frac{1}{2} \tr \sqrt{ ( \rho_1 - \rho_2 )^2 }$
denotes the trace distance between two density matrices \cite{degenerate}.
This is a natural distance measure on
density matrices, giving the maximum difference in probability
for any measurement on the two states \cite{norms}.
The meaning of eq.~(\ref{prev_result}) is that the average distance
between the instantaneous state of the subsystem
$\rho_S(t)$ and the fixed state $\omega_S$ will be small whenever the total
dimension explored by the state (or
the dimension explored in the bath) is much larger than the subsystem dimension.
In typical thermodynamic situations, this will indeed be the case. Indeed, as
dimensions typically grow exponentially with particle
number, we would expect any expression of the form
$\mathrm{poly}(d_S)/d_{\eff}(\omega)$ to tend to zero in the
thermodynamic limit (where the fraction of particles in the system tends
to zero), as long as the energy distribution of $\ket{\Psi(0)}$ was reasonable.
This addresses point 1 in our programme;
in practice one has to check that an initial state leads to large enough
$d^{\eff}(\omega)$; however, \cite[Theorem 2]{linden} shows that this is the case
for typical states from any sufficiently large subspace (for instance,
a subspace of states with similar macroscopic properties).

Regarding item {\it 2} it is proven \cite[Theorem 3]{linden} that
initial states of tensor product form of a fixed state on the
subsystem with a typical state of a large enough subspace on the bath,
yield very similar equilibrium states.

Finally, regarding item {\it 3},
we similarly prove \cite[Theorem 3]{linden} that if the energy
eigenstates of the Hamiltonian are sufficiently entangled, then also
initial states of tensor product form of a typical state on the system
with a fixed state on the bath have very similar equilibrium states.
Furthermore, simple examples such as Hamiltonians with tensor product
eigenstates, show that without any additional assumptions on the
correlation properties of the eigenstates one cannot expect
subsystem independence to hold.

We emphasize as in \cite{linden} that in the above discussion the
``equilibrium'' state $\omega_S$ is not necessarily Boltzmannian, and
may depend on details of the Hamiltonian and the initial state (in
particular, strong interactions, or conserved quantities for the
subsystem will generate different equilibrium states). However, the
equilibration results still hold.

We now come to a crucial further issue, namely time scales.  This issue needs to be a part of the general
programme of investigating thermalisation. Again, this issue can be decomposed in a number of different
questions. Firstly ''How long does it take for a system to come close to equilibrium?'', and secondly ''What is
the speed of fluctuations around equilibrium?''. This second point is what concerns us here.

While the magnitude of the fluctuations from equilibrium may
be small according to eq.~(\ref{prev_result}),
this does not say anything about their speed.
Mathematically, the speed of change of the subsystem state is given by
\begin{equation} \label{vel_def}
v_S(t) = \lim_{\delta t \rightarrow 0} \frac{D(\rho_S(t),
\rho_S(t+\delta t) )}{\delta t}     = \frac{1}{2} \left\| \frac{d
\rho_S(t)}{d t} \right\|_1.
\end{equation}
We will show that $v_S(t)$ is small for almost all times in typical
thermodynamic cases, as follows.

\bigskip
\noindent
{\bf Theorem.} For the average rate of change of $\rho_S$ it holds that
\begin{equation}
\av{v_S(t)}\leq   \|H_S  +H_{\rm \rm int}\| \sqrt{ \frac{
d_S^3}{d_{\eff} (\omega )}},\label{theorem}
\end{equation}
where we take $\hbar=1$, and use the operator norm \cite{norms}.

\medskip
\noindent
\textit{Proof.}
The time evolution of the subsystem state is given by
\begin{equation}
\frac{d \rho_S(t)}{d t}  =\tr_B ( i [\rho(t), H]) = \sum_k c_k (t) e_k,
\end{equation}
where $e_k$ with $k =\{1,2,..., d_S^2\}$ is a Hermitian orthonormal
operator basis for the system, i.e.~$\tr (e_k e_\ell) = \delta_{k\ell}$.
Hence,
\begin{eqnarray}
c_k (t) &=& \tr_S \left( \frac{d \rho_S(t)}{d t} e_k \right) \nn \\
    &=& \tr \left( i \left[\rho(t), H \right] e_k \otimes I \right) \nn \\
    &=& \tr \left( \rho(t) i[H, e_k \otimes I] \right) \nn \\
    &=& \tr \left(  \rho(t) i[H_S +H_{\rm int} , e_k \otimes I] \right).
\label{evolution}
\end{eqnarray}

Using our notation, and with a slight modification to use the
operator norm, Reimann \cite{reimann} shows that for a Hamiltonian
with non-degenerate energy gaps, and a Hermitian observable $A$,
\begin{equation}
\av{\left( \tr(\rho (t) A) -  \av{\tr(\rho(t) A) } \right)^2}\leq
\frac{\|A\|^2}{d_{\eff} (\omega )}.
\end{equation}
Taking $A=i[H_S+H_{\rm int} , e_k \otimes I]$, and noting that
\begin{equation}
  \left\| i[H_S +H_{\rm int} , e_k \otimes I] \right\|
      \leq 2 \|H_S + H_{\rm int}\|,
\end{equation}
we obtain
\begin{equation}
\av{\left( c_k (t) - \av{c_k (t)} \right)^2 }\leq \frac{4 \|H_S +
H_{\rm int}\|^2}{d_{\eff} (\omega )}.
\end{equation}
However,
\begin{equation}\begin{split}
  \av{c_k (t)} &= \av{\tr \left( i \left[\rho(t), H \right] e_k \otimes I \right)}  \\
               &= \tr \left( i \left[\omega, H \right] e_k \otimes I \right) = 0,
\end{split}\end{equation}
hence
\begin{equation}
\av{\left( c_k (t)  \right)^2 }\leq \frac{4 \|H_S + H_{\rm
int}\|^2}{d_{\eff} (\omega )}.
\end{equation}
This implies
\begin{eqnarray}
\av{\left\| \frac{d \rho_S(t)}{d t} \right\|_2^2}  &=& \av{ \tr \left( \sum_k c_k (t) e_k \right)^2} \nn \\
    &=&  \sum_{kl} \av{ c_k (t) c_l(t) }  \tr \left( e_k e_l \right) \nn \\
    &=&  \sum_k \av{(c_k (t))^2} \nn  \\
    &\leq&  \frac{ 4 \|H_S + H_{\rm int}\|^2d_S^2}{d_{\eff} (\omega )}.
\end{eqnarray}
On the other hand, the trace norm and Hilbert-Schmidt norm
are connected by the elementary relation
$\| X \|_1^2 \leq ({\operatorname{rank} X}) \| X \|_2^2$. Hence, and
using the concavity of the square root function,
\begin{eqnarray}
\av{ \left\| \frac{d \rho_S(t)}{d t} \right\|_1 } &\leq& \sqrt{\av{ \left\| \frac{d \rho_S(t)}{d t} \right\|_1^2 }} \nn \\
    &\leq& \sqrt{d_S \av{  \left\| \frac{d \rho_S(t)}{d t} \right\|_2^2 }} \nn \\
    &\leq& 2 \|H_S + H_{\rm
int}\| \sqrt{ \frac{ d_S^3}{d_{\eff} (\omega )}}.
\end{eqnarray}
From the definition of $v_S(t)$ given by (\ref{vel_def}), we obtain
the desired result
\begin{equation}\begin{split}
  \av{v_S(t)} &= \frac{1}{2} \av{ \left\| \frac{d \rho_S(t)}{d t} \right\|_1 } \\
              &\leq  \|H_S + H_{\rm int}\| \sqrt{ \frac{ d_S^3}{d_{\eff} (\omega )}}.
  \quad\quad\quad\quad \qed
\end{split}\end{equation}

\medskip
This result can be interpreted as follows. First, the speed of
fluctuations varies in time and, of course, there may -- and
in general will -- be times
when the speed is extremely high. What our theorem shows is
that, on average, the instantaneous speed is bounded by the
expression given in eq. (\ref{theorem}). Since speed is a positive
quantity, this also means that the fraction of times for which $v_S(t)
> K \|H_S + H_{\rm int}\| \sqrt{ \frac{ d_S^3}{d_{\eff} (\omega )}}$
must be less than $1/K$.

Second, the speed of fluctuations depends linearly on the magnitude
of the Hamiltonian, more precisely, on the subsystem and interaction
Hamiltonian. Clearly the speed needs to depend linearly on the
Hamiltonian since multiplying the Hamiltonian by a constant factor
$H\rightarrow \lambda H$ makes the entire time evolution faster by
the factor $\lambda$.  Furthermore, the speed of change of the state
of the subsystem depends only on the part of the Hamiltonian
that acts directly on it, in particular it is independent of the bath
Hamiltonian $H_B$. Of course, since the subsystem interacts with
the bath, the time evolution of the subsystem depends on the state
of the bath and thus implicitly on the bath Hamiltonian. However, as
is already evident in eq. (\ref{evolution}), the instantaneous change
in the state of the subsystem (and hence the speed of its evolution)
depends only on the instantaneous state of he bath and not directly
on the bath Hamiltonian. Also, obviously, the time evolution is
completely independent on the constant part of the Hamiltonian,
$H_0$ that only defines a reference point for the energy. These
being said, the bound (\ref{theorem}) should be better interpreted
as a bound on the speed of fluctuations as measured in ``natural
units'' i.e.
\begin{equation}
{{\av{v_S(t)}}\over{ \|H_S+H_{\rm \rm int}\| }} \leq  \sqrt{ \frac{
d_S^3}{d_{\eff} (\omega )}}.\label{theorem2}
\end{equation}

The main result of this  paper is therefore that the average speed,
as measured in natural units, is bounded by $\sqrt{ \frac{
d_S^3}{d_{\eff} (\omega )}}$. As mentioned earlier, because
dimension generally grows exponentially with particle number, we
would expect any fixed power of $d_S$ to be much smaller than
${d_{\eff} (\omega )}$ in the thermodynamic limit.

Note that similar arguments can be made for any finite derivative
of $\rho_S(t)$, exhibiting higher powers of the Hamiltonian
in the upper bound with increasing degree of the time derivative.

Finally, as far as the absolute value of the speed is concerned, we
note that in most natural systems the magnitude of the Hamiltonian
governing the speed, i.e. $\|H_S+H_{\rm \rm int}\| $ scales
relatively slowly (i.e. polynomially) with the number of particles.
For example, in a system of $n$ particles in which the Hamiltonian
only contains two-particle interactions, we would expect the norm of
the Hamiltonian to grow at most quadratically in $n$. Hence in the
thermodynamic limit, when the total number of particles in the
system increases, we would expect the exponential dependence of the
dimensional term  $\sqrt{ \frac{d_S^3}{d_{\eff}(\omega )}}$ to
dominate, causing the absolute value of the average speed to tend to
zero.

To summarize, we have shown that in the thermodynamic limit not only
every subsystem spends almost all its
time fluctuating very closely around a fixed state -- the equilibrium
state -- but also that the speed of
fluctuations becomes vanishingly small. Both these results at first
glance appear to contradict the established view
that a system has non-vanishing fluctuations around equilibrium.
However, as we have already argued in \cite{linden}, the
global state $\rho_S(t)$ of a subsystem fluctuates extremely little around
the equilibrium state, and what are
generally thought to be time fluctuations are just the result of quantum
uncertainty (the probabilistic nature of the outcomes of quantum measurements),
that would be present even if the state were absolutely time independent.

As in \cite{linden}, our results hold not only for the
standard statistical mechanical setting with Boltzmannian equilibrium,
but under extremely general conditions.

\bigskip
\noindent
\textit{Acknowledgments.}
This research was supported by the ``QIP IRC'' of the U.K.~EPSRC
and the IP ``QAP'' of the European Commission.
AJS is supported by the Royal Society.
AW is supported by the U.K.~EPSRC, the Royal Society, and a
Philip Leverhulme Prize.
The Centre for Quantum Technologies is funded by the
Singapore Ministry of Education and the National Research Foundation
as part of the Research Centres of Excellence programme.


\begin{thebibliography}{10}
\bibitem{norms} We will make use of several norms on Hermitian operators,
    in particular the trace
    norm $\|A\|_1 = \tr|A| = \sum_i |a_i|$, where $a_i$ are the eigenvalues of $A$;
    the Hilbert-Schmidt norm $\|A\|_2=\sqrt{\tr(A^{\dag}A)} = \sqrt{\sum_i |a_i|^2}$;
    and the standard operator norm  $\|A\| = \max_i |a_i|$.
\end{thebibliography}
\end{document}